\documentclass[pra,twocolumn,showpacs]{revtex4}
\usepackage{epsfig}
\usepackage{graphicx}
\usepackage{amsmath}
\usepackage{natbib}

\newcommand{\n}{\nonumber}
\newcommand{\bn}{\begin{eqnarray}}
\newcommand{\en}{\end{eqnarray}}
\newcommand{\eml}{\end{multline}}
\newcommand{\bml}{\begin{multline}}

\newcommand{\op}[1]{\hat{#1}}

\begin{document}

\title {Spin squeezing by tensor twisting and Lipkin-Meshkov-Glick dynamics in a toroidal Bose-Einstein condensate with spatially modulated nonlinearity}
 \author{Tom\'{a}\v{s} Opatrn\'{y}$^1$, Michal Kol\'{a}\v{r}$^1$, and Kunal K. Das$^2$}
 \affiliation{$^1$Optics Department, Faculty of Science, Palack\'{y} University, 17. Listopadu 12,
 77146 Olomouc, Czech Republic\\
 $^2$Department of Physical Sciences, Kutztown University of Pennsylvania, Kutztown, Pennsylvania 19530, USA}

\date{\today }
\begin{abstract}
We propose a scheme for spin-squeezing in the orbital motion of a Bose-Einstein condensate (BEC) in a toroidal trap. A circular lattice couples two counter-rotating modes and squeezing is generated by the nonlinear interaction spatially modulated at half the lattice period. By varying the amplitude and phase of the modulation, various cases of the twisting tensor can be directly realized, leading to different squeezing regimes. These include one-axis twisting and the two-axis counter-twisting which are often discussed as the most important paradigms for spin squeezing. Our scheme naturally realizes the Lipkin-Meshkov-Glick model with the freedom to vary all its parameters simultaneously.
\end{abstract}

\maketitle

\section{Introduction}

Squeezing in an ensemble of two-level systems \cite{Kitagawa,Wineland1994} is a quantum phenomenon with applications ranging from quantum information \cite{Memory} to precision metrology  \cite{Wineland1994,Lloyd}. Since the canonical two-level system is a spin-$\frac{1}{2}$ particle, this is often referred to as ``spin squeezing'', but physical realizations include a broad range of systems defined by the same Lie algebra, such as two-component BECs \cite{Esteve2008,BosonicJosephson} or polarized light \cite{Korolkova}. A seminal paper by Kitagawa and Ueda \cite{Kitagawa} established nonlinear dynamics as a natural way to generate spin-squeezing,  via two distinct mechanisms: by one-axis twisting (OAT) on the Bloch sphere with Hamiltonian $\sim \hat{J}_z^2$, and two-axis counter-twisting (TACT) with Hamiltonian $\sim \hat{J}_x^2-\hat{J}_y^2$, with $\hat{J}_{x,y,z}$ being components of the collective spin operator. The latter scenario is more efficient in generating strong spin squeezing, however no experiment has achieved it directly, but schemes have been proposed to convert OAT into effective TACT Hamiltonians \cite{Liu2011}.

\begin{figure}[t]
\centerline{\epsfig{file=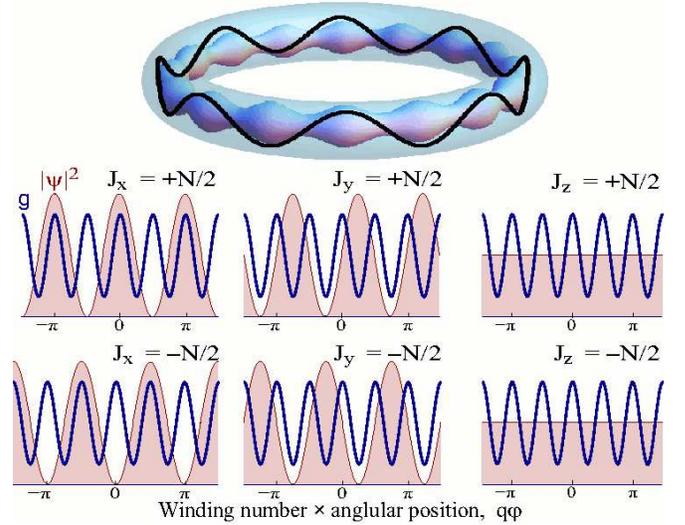,width=\columnwidth}}
\caption{\label{f-schematic} (Color online) Schematic of our model of a BEC in a toroidal trap with a circular lattice (black line) and nonlinearity (inner torus) modulated at half the lattice period. Plots show relative positions of the density of the BEC $|\Psi(\varphi)|^2$ (shaded area) and of the nonlinear strength $g(\varphi)$ (solid line) modulated at half the spatial period for states with extremal $\langle\hat{J}_{x,y,z}\rangle$, shown here for the case of TACT with $g_1=\frac{2}{3}g_0$.}
\end{figure}

The spin-squeezing Hamiltonian quadratic in $\hat{J}_{x,y,z}$ is part of the more general Lipkin-Meshkov-Glick (LMG) model \cite{Lipkin}, an important exactly solvable model
introduced to test various many-body approximation methods. The model is relevant in several fields, including nuclear physics \cite{Lipkin,Narducci-nuclear}, quantum criticality and phase transitions \cite{LMG-quantum-phase}, molecular magnetism \cite{Chudnovsky2001}, multi-particle entanglement \cite{Vidal-entanglement} and classical bifurcations \cite{Zibold2010, Castanos2006}. The broad applicability of the LMG Hamiltonian,
\begin{eqnarray}
\hat{H}_{\rm LMG}=\hbar[\Omega \hat{J}_z + W (\hat{J}_x^2 + \hat{J}_y^2) + V (\hat{J}_x^2 - \hat{J}_y^2)],
\end{eqnarray}
arises from the independent variation of the three parameters $\Omega, W, V$.  Therefore, ultracold atomic systems, with their extreme tunability and control, seem a natural choice for a LMG simulator. But as yet only a limited case with one varied parameter (keeping $W=V$ fixed) has been demonstrated with BEC \cite{Zibold2010}.

In this paper, we propose a way to implement general LMG with independent multi-parameter control as well as direct spin squeezing by both TACT and OAT, with a BEC in a ring trap with periodic angular modulation of the external potential and of the inter-particle interactions \cite{Yamazaki, Yulin}. Two counter-propagating rotational modes act as pseudo-spin coupled by the periodic potential; and the spatially modulated nonlinearity controls and varies the various components of the twisting tensor $\chi$ in the general quadratic Hamiltonian
\begin{eqnarray}
\hat{H}=\hbar \left[
\sum_k \omega_k \hat{J}_k +\sum_{kl} \chi_{kl}\hat{J}_k \hat{J}_l\right]
\end{eqnarray}
that covers OAT, TACT, and LMG as special cases \cite{TO2014}.

Our study is also motivated by the rapid advances in recent years in toroidal trapping of ultracold atoms by a variety of ways: by intersecting a ``sheet" beam with a Lagueree-Gaussian beam \cite{Ramanathan} or a beam passed through a ring-shaped mask \cite{Eckel,Jendrzejewski}, with painted potentials \cite{Ryu}, and with magnetic waveguides \cite{Gupta2005}. Notably, the study of LMG and spin-squeezing provides a distinct new line of study with such traps which have so far been primarily used to examine superfluid flow \cite{Ramanathan, Eckel,Jendrzejewski} and interferometric effects \cite{Ryu}.

The paper is organized as follows. In Section \ref{derivation} we derive the Hamiltonian, in Sec. \ref{spinsqueezing} spin squeezing by OAT and TACT is discussed, in Sec \ref{LMGmodel} we discuss the LMG model and phase transitions in our scheme, in Sec. \ref{fastestsqueezing} we show various features of the spin squeezing procedure under the optimum rotation condition, in Sec. \ref{conditions} we specify the conditions for physical parameters, in Sec. \ref{stateprep}
we discuss various options for initial state preparation and detection, and we conclude with Sec. \ref{conclusion}. Some mathematical details are in Appendices.


\section{Derivation of the Hamiltonian}
\label{derivation}

Consider a BEC of $N$ atoms  in a toroidal trap with a superimposed potential ${\cal U}$ and spatially varying nonlinearity, with the latter having twice the periodicity of the potential around the ring as shown in Fig.~\ref{f-schematic}. For example, in a Laguerre-Gaussian implementation this could be achieved by modulating the amplitudes and phases of the component beams. The spatially varying nonlinearity can be produced by optically induced Feshbach resonance \cite{Theis2004}, as demonstrated for linear geometry in a recent experiment \cite{Yamazaki}. The dynamics is driven by the Hamiltonian  \cite{pitaevski-stringari}
\begin{eqnarray}
&&\op{H}\!=\! \int{\rm d}\mathbf{r}\op{\Psi}^\dagger(\mathbf{r}) \left[- \frac{\hbar^2}{2m}\triangle+{\cal U}(\mathbf{r},t)+\frac{g(\mathbf{r},t)}{2} \op{\Psi}^\dagger\op{\Psi}\right] \op{\Psi}(\mathbf{r}),\n\\
&&{\rm where}\ \
g(\mathbf{r},t)= \frac{4\pi \hbar^2 a (\mathbf{r},t)}{m}
\label{QF-Hamiltonian}\end{eqnarray}
is the nonlinearity parameter,
with $a $ being the scattering length, $m$  the particle mass, and $\op{\Psi}(\mathbf{r })$, $\op{\Psi}^\dagger(\mathbf{r^\prime})$ are, respectively the annihilation and creation bosonic field operators satisfying the commutation relations $[\op{\Psi}(\mathbf{r}),\op{\Psi}^\dagger(\mathbf{r} ^\prime)]=\delta(\mathbf{r}-\mathbf{r^\prime})$.

Using cylindrical co-ordinates ${\mathbf r}=(r,z,\varphi)$,
with $z$ along ring axis, we assume tight harmonic confinement in radial and axial directions
\begin{eqnarray}
{\cal U}(\mathbf{r},t)=\frac{1}{2}m\omega_z^2 z^2+\frac{1}{2}m\omega_r^2 (r-R)^2+U(\varphi,t)
\end{eqnarray}
with $R$ being mean ring radius. The azimuthal potential is taken to be weakly sinusoidal rotating along the ring with frequency $\omega$, creating a circular Bragg grating,
\begin{eqnarray}
U(\varphi,t) = \hbar u_x \cos\left[2q (\varphi + \omega t) \right]
+ \hbar u_y \sin\left[2q (\varphi + \omega t) \right] ,
\end{eqnarray}
with $q$ an integer and amplitudes assumed small, $u_{x,y} \ll \omega_{z,r}$. The nonlinear parameter is given a periodic spatio-temporal dependence,
\begin{eqnarray}
g(\mathbf{r},t)=g(\varphi,t) = g_0 + g_1 \cos \left[4q (\varphi + \omega t) -\alpha \right],
\end{eqnarray}
where $g_0$, $g_1$ and $\alpha$ are constants.

We expand the field operators $\op{\Psi}(\mathbf{r})=\sum_n \hat{a}_n\psi_n({\mathbf r})$  in terms of the mode functions of the ring trap,
\begin{eqnarray}&&\psi_{n}(\mathbf{r}) = \frac{1}{2\pi R\sigma_z \sigma_R }
e^{in\varphi}  \times e^{(r-R)^2/(4\sigma_R^{2})} \times
e^{- z^2/(4\sigma_z^{2})}\n\\&&{\rm with}\ \
\sigma_{z,R} = \sqrt{\frac{\hbar}{2m\omega_{z,R}}}
\end{eqnarray}
being the rms widths of the mode functions in the respective directions. Assuming that the tight trapping allows only the ground state to be populated in the axial and radial directions, the Hamiltonian can be written in
terms of the annihilation and creation operators $\op{a}_{n}$ and $\op{a}_{n}^{\dag}$ as
\begin{widetext}
\begin{eqnarray}
\label{Hama}
 \hat{H} = \frac{\hbar^2}{2mR^2}\sum_{n} n^2\op{a}_{n}^{\dag}\op{a}_{n}  +  \frac{\hbar u_x}{2}\sum_{n} \left( \op{a}_{n}^{\dag}\op{a}_{n-2q} e^{i2q\omega t} + \op{a}_{n}^{\dag}\op{a}_{n+2q} e^{-i2q\omega t} \right) +  \frac{\hbar u_y}{2i}\sum_{n} \left( \op{a}_{n}^{\dag}\op{a}_{n-2q} e^{i2q\omega t} - \op{a}_{n}^{\dag}\op{a}_{n+2q} e^{-i2q\omega t} \right) \nonumber \\
  + \frac{1}{16 \pi^2 \sigma_R \sigma_z R} \sum_{n,k,l,s}
\op{a}_{n}^{\dag}\op{a}_{k}^{\dag} \op{a}_{l} \op{a}_{s} \left[ g_0 \delta_{n+k,l+s}  +  \frac{g_1  }{2}\left(\delta_{n+k,l+s+4q} e^{i(4q\omega t-\alpha)}+
\delta_{n+k,l+s-4q} e^{-i(4q\omega t-\alpha)}
\right) \right].
\end{eqnarray}
\end{widetext}

The Hamiltonian simplifies further if only two counter-propagating modes with $n=\pm q$ are initially populated, and the energy gap manifest in the diagonal terms inhibits transitions to other modes.
We absorb the time dependence of the fields into the definition of the mode operators,
$\op{a}_{q} \equiv \op{a} e^{iq\omega t} $,
$\op{a}_{-q} \equiv \op{b} e^{-iq\omega t}$,
and we define the pseudo-spin operators $\op{J}_{x,y,z}$
\begin{eqnarray}
\label{Jx}
\op{J}_x &=& \frac{1}{2}(\op{a}^{\dag}\op{b}+\op{a}\op{b}^{\dag}), \n\\
\op{J}_y &=& \frac{1}{2i}(\op{a}^{\dag}\op{b}-\op{a}\op{b}^{\dag}), \n\\
\op{J}_z &=& \frac{1}{2}(\op{a}^{\dag}\op{a}-\op{b}^{\dag}\op{b}).
\label{Jz}
\end{eqnarray}
the Hamiltonian reduces to
\begin{widetext}
\begin{eqnarray}
\hat{H} = \hbar \left[u_x \op{J}_x + u_y \op{J}_y  + 2q\omega \op{J}_z - \chi_0 \op{J}_z^2  +\frac{\chi_1 \cos \alpha }{2}\left(\op{J}_x^2 - \op{J}_y^2 \right)
+\frac{\chi_1 \sin \alpha }{2}\left(\op{J}_x \op{J}_y + \op{J}_y \op{J}_x \right)
\right]
\label{HamJ}
\end{eqnarray}
\end{widetext}
where we have defined effective one-dimensional nonlinear coefficients
\begin{eqnarray}
\chi_{0,1} \equiv  \frac{g_{0,1}}{8 \pi^2 \hbar \sigma_R \sigma_z R}
=  a _{0,1}\frac{ \sqrt{\omega_R \omega_z}}{\pi R} ,
\end{eqnarray}
and  dropping a scalar term $\chi_0 \left(\frac{3}{4}N^2 -\frac{1}{2}N \right)$ that creates an irrelevant global phase. This effective Hamiltonian is one of main result of the paper.  Additional details of its derivation is provided in Appendix \ref{appendixA}.

It is worth noting that although the sign of the unmodulated nonlinear term $-\hbar\chi_0\hat{J}_z^2$ suggests attractive inter-particle interaction, it actually
stems from repulsive interaction for positive $g_0$. This apparent contradiction stems from the fact that for $J_z\approx 0$ the counter-rotating modes $\op{a}$ and  $\op{b}$ are almost equally populated and the condensate forms a standing wave with pronounced interference fringes (see Fig. \ref{f-schematic}). Therefore, the particles are effectively compressed to half of the volume that would be occupied if they were all circulating in the same direction ($J_z\approx \pm N/2$)
with no interference fringes. Thus, states with larger $|J_z|$ correspond to lower interaction energy.

\section{Spin-squeezing by OAT and TACT}
\label{spinsqueezing}

The terms linear in $\op{J}_{x,y,z}$ generate rotation on the Bloch sphere with angular velocity vector $(u_x,u_y,2q \omega)$. Spin squeezing is generated by nonlinear terms the coefficients of which can be written as a \emph{twisting} tensor \cite{TO2014}:
\begin{eqnarray}
\chi = \left(
\begin{array}{ccc}
\frac{\chi_1}{2}\cos \alpha & \frac{\chi_1}{2}\sin \alpha & 0  \\
\frac{\chi_1 }{2}\sin \alpha  & -\frac{\chi_1}{2}\cos \alpha & 0 \\
0 & 0 & - \chi_0
\end{array}
\right)  .
\label{twist}
\end{eqnarray}
Since $\sum_k \op{J}_k \op{J}_k =\frac{N}{2}\left(\frac{N}{2} +1\right)$, adding multiples of a unit matrix to $\chi$ only leads to a trivial global phase. For $\chi_1=0$ (no spatial modulation of the nonlinearity) the position of the BEC interference fringes plays no role. The Hamiltonian is invariant with respect to rotations around $J_z$, and the nonlinear dynamics corresponds to OAT. When  $\chi_1\neq 0$  the nonlinearity is modulated, and the interaction energy is affected by the relative positions of the interference fringes and the maxima of the nonlinearity (see Fig. \ref{f-schematic}). In this case the twisting tensor has eigenvalues $-\chi_0$ and  $\pm |\chi_1|/2$.  The maximum squeezing rate is determined by the difference between the largest and the smallest eigenvalues of the twisting tensor \cite{TO2014}. For positive $\chi_{0}$ and states that are nearly Gaussian and optimally positioned on the Bloch sphere, on the axis of the middle eigenvalue, the maximum squeezing rate is
\begin{eqnarray}
\frac{d\xi^2}{dt} = -N\left( \chi_0 + \frac{|\chi_1|}{2}\right) \xi^2,
\end{eqnarray}
where the squeezing parameter $\xi^2$ is defined as the ratio of the minimum variance of the uncertainty ellipse of the state and the variance of the spin coherent state. Appendix \ref{appendixB} provides more details about the squeezing.
To achieve this maximum rate, the state should be centered on the Bloch sphere along the axis corresponding to the middle eigenvalue, and kept optimally oriented by continuous rotation of the system at frequency  $N\left(\frac{\chi_a+\chi_c}{2} -\chi_b \right)$ around this axis, where $\chi_{a,b,c}$ are the eigenvalues in ascending order; here, if $|\chi_1| \leq \chi_0$,  this frequency is $N\left(\frac{3}{4}|\chi_1|-\frac{1}{2}\chi_0\right)$.

In the special case when the middle eigenvalue is exactly between the other two eigenvalues, $\chi$ can be transformed to a diagonal form diag$(\tilde{\chi},-\tilde{\chi},0)$ corresponding to TACT (see \cite{TO2014} for details); here $2\tilde{\chi}$ is the difference between the largest and smallest eigenvalues.  In our model it corresponds to the condition $|\chi_1| = \frac{2}{3}\chi_0$.
Since the modulation maxima and minima are proportional to $\chi_0\pm |\chi_1|$, the maximum and minimum are $\frac{5}{3}\chi_0$ and  $\frac{1}{3}\chi_0$, so that
TACT occurs if the maximum value of the modulated nonlinearity is five times larger than the minimum (see Fig. \ref{f-schematic}). In this case, the state naturally remains optimally oriented, no rotation is needed.

\section{The LMG Model and Phase transitions}
\label{LMGmodel}
Special cases of our Hamiltonian (\ref{HamJ}) correspond to
various instances of the LMG model. For example, setting $u_{x,y}=\alpha=0$ leads to LMG model with $\Omega=2q\omega$, $W=\chi_0$,
and $V=\chi_1/2$, and setting $u_{y}=\omega=\alpha=0$ leads to LMG model with $\Omega=u_x$, $W=-\frac{1}{2}\chi_0-\frac{3}{4}\chi_1$, and  $V=\frac{1}{2}\chi_0-\frac{1}{4}\chi_1$  with the roles of the coordinates in $\op{H}_{LMG}$ permutated $J_x\to J_y \to J_z \to J_x$.

\begin{figure}
\centerline{\epsfig{file=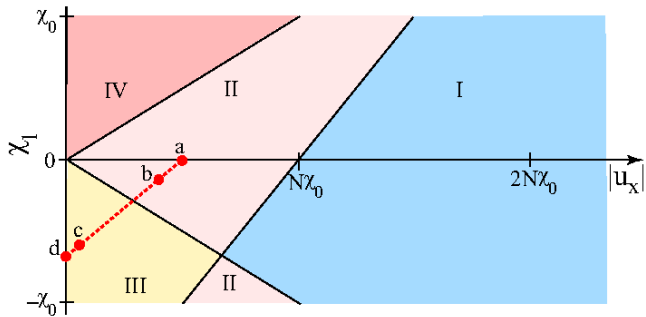,width=\columnwidth}}
\caption{\label{f-phasediag1} (Color online) LMG phase diagram in the $u_x$---$\chi_1$ plane with fixed $\chi_0$, and $u_y = \omega=\alpha=0$. The phases I---IV  differ in the number and type of singularities in the density of states, and correspond to those of Ref.~\cite{Ribeiro2007}. The dashed red line corresponds to optimum rotation $u_x$ at given $\chi_{0,1}$ for fastest squeezing, with points $a\rightarrow$ OAT, $\chi_1=0$; $b\rightarrow \chi_1=-0.2 \frac{2}{3}\chi_0$; $c\rightarrow \chi_1=-0.9 \frac{2}{3}\chi_0$; $d\rightarrow$  TACT, $\chi_1= -\frac{2}{3}\chi_0$. Time evolution of squeezing at these points are in Fig.~\ref{f-timevol1}.}
\end{figure}

To illustrate the applicability of our Hamiltonian to the LMG model, we show a possible LMG phase diagram in Fig. \ref{f-phasediag1}. In this example, we assume $u_y = \omega=\alpha=0$ and $\chi_0>0$, varying $\chi_1 \in (-\chi_0,\chi_0)$ and $u_x$. Such a choice may be natural, for instance, if one needs to avoid rotation of the fields.  The zones denoted I---IV correspond to those studied in \cite{Ribeiro2007}, reflecting different qualitative behavior of the energy surfaces and different properties of the density of eigenstates of the Hamiltonian. Crossing the boundary between various zones corresponds to various phase transitions; for example crossing from  I to II by decreasing $|u_x|$ with $\chi_1=0$ corresponds to the classical bifurcation at the transition from Rabi to Josephson dynamics observed in \cite{Zibold2010}.

\subsection{Fastest squeezing within the LMG model}
\label{fastestsqueezing}

The conditions for fastest squeezing can be achieved with the parameters used for the LMG phase diagram above. The state is positioned along $J_x$ axis
on the Bloch sphere and rotation $u_x$ should be applied at the optimum frequency defined earlier. These conditions are fulfilled by parameters of the red broken line between points a---d in Fig. \ref{f-phasediag1}. We show the corresponding time evolution of the squeezing parameter $\xi^2$ in Fig. \ref{f-timevol1}. The best squeezing properties correspond to TACT (line d) and as can be seen, even a relatively small deviation of the TACT condition leads to a strong deterioration of squeezing in the last stages (line c).

\subsection{Energy spectrum and energy eigenstates}

For the respective parameter values we also show the density of states $\rho(\epsilon)$ as a function of the Hamiltonian eigenvalues $\epsilon$ with qualitatively different shapes in zone II (one peak) and III (two peaks that merge in the limit of TACT, point d of Fig. \ref{f-phasediag1}). The peaks divide the energy spectrum into regions with qualitatively different features of the Hamiltonian eigenstates. As an interesting relationship between squeezing and LMG phase we note that the states undergoing squeezing with the optimum rotation are located at one of the density peaks.

\begin{figure}[t]
\centerline{\epsfig{file=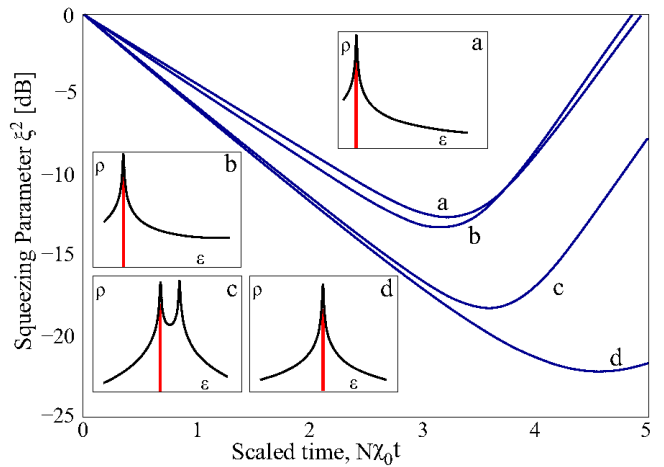,width=\columnwidth}}
\caption{\label{f-timevol1} (Color online) Time evolution of the squeezing parameter $\xi^2$ with $N=300$ and optimum rotation $u_x$, with different values of $\chi_1$ marked in Fig.~\ref{f-phasediag1}:
$a\rightarrow$ OAT, $\chi_1=0$; $b\rightarrow \chi_1=-0.2 \frac{2}{3}\chi_0$; $c\rightarrow \chi_1=-0.9 \frac{2}{3}\chi_0$; $d\rightarrow$  TACT, $\chi_1= -\frac{2}{3}\chi_0$. Insets show the corresponding density of states $\rho$ as function of the energy spectrum $\epsilon$ of the Hamiltonian; position and width of vertical red bars show mean and spread of energy of the states used for the curves in the main plot.}
\end{figure}

\begin{figure}[t]
\centerline{\epsfig{file=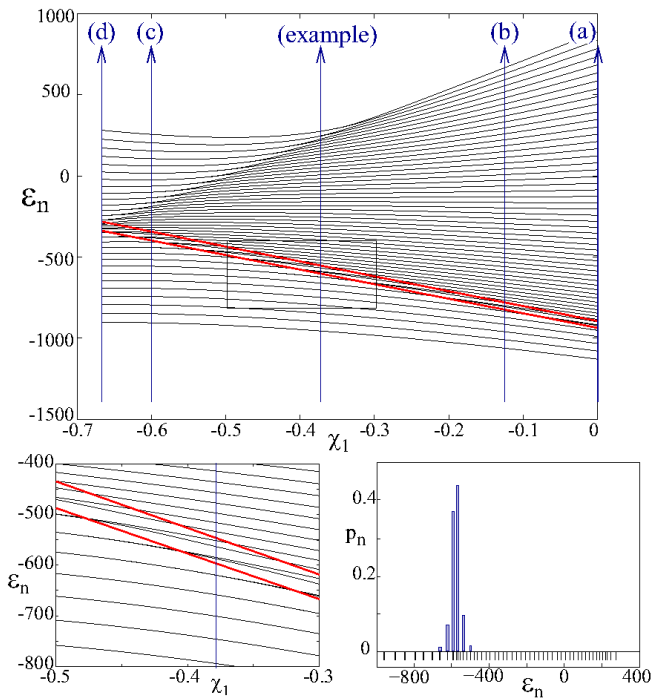,width=\columnwidth}}
\caption{\label{f-eigenenergs1} (Color online)  Eigenvalues of the Hamiltonian with $N=60$, $\chi_0=1$, as functions of $\chi_1$, with $u_x$ chosen to be $u_x =N\left(\frac{3}{4}|\chi_1|-\frac{1}{2}\chi_0\right)$ to achieve fastest squeezing. The vertical arrows labeled (a)---(d) correspond to the cases shown in Figs. \ref{f-phasediag1} and \ref{f-timevol1}.  The arrow indicated by
`(example)' corresponds to the case of the  $Q$-functions in Fig.~\ref{f-Qfunctions1}  with
$\chi_1=-0.373$.  The two transverse red lines show the energy spread $\bar E \pm \Delta E$ of the state being squeezed. Details of the region marked by the rectangle are shown in the lower left panel. The histogram shows the probability distribution of the Hamiltonian eigenstates that constitute the state being squeezed, marked by the line labeled `(example)'. The ticks below the axis show the positions of the eigenvalues.}
\end{figure}

\begin{figure}[t]
\centerline{\epsfig{file=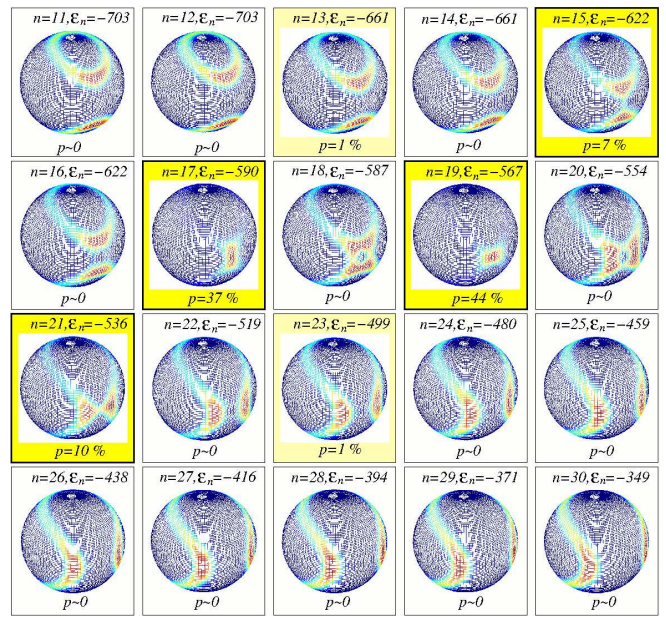,width=\columnwidth}}\vspace{-2mm}
\caption{\label{f-Qfunctions1} (Color online) $Q$-functions of different eigenstates $\psi_n$ of the Hamiltonian with $n=16, \dots, 30$, shown for specific $\chi_1=-0.373$ marked as `(example)' in Fig.~\ref{f-eigenenergs1}, with the other parameters the same, $N=60$, $\chi_0=1$, $|u_x|=0.22 N$. In each panel the energy $\epsilon_n$ of the state is indicated, as well as the probability $p$ of this state to be found in the spin coherent state of $J_x=N/2$. The highlighted panels show the most important constituent eigenstates of the state being squeezed.}
\end{figure}

\begin{figure}[t]
\centerline{\epsfig{file=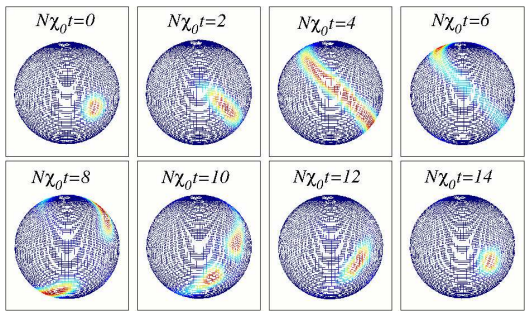,width=0.8\columnwidth}}\vspace{-2mm}
\caption{\label{f-Qfunctions2} (Color online) Time evolution of the $Q$-function  with $N=60$, $\chi_0=1$, $\chi_1=-0.373$, $u_x =N\left(\frac{3}{4}|\chi_1|-\frac{1}{2}\chi_0\right)$
of an initially spin coherent state. This state is formed as a superposition
of the eigenstates shown in Fig. \ref{f-Qfunctions1}. }
\end{figure}

We illustrate this feature in a few figures. In Fig.~\ref{f-eigenenergs1} we show the eigenvalues for the Hamiltonian from  Eq.~(\ref{HamJ}) as a function of $\chi_1$ with the remaining parameters fixed at $N=60$, $\chi_0=1$, $u_y=\omega = \alpha =0$, and choosing  $u_x =N\left(\frac{3}{4}|\chi_1|-\frac{1}{2}\chi_0\right)$ for fastest squeezing. We set $\hbar=1$.
The arrow denoted ``(example)'' corresponds to the Bloch sphere diagrams in Figs.~\ref{f-Qfunctions1} and \ref{f-Qfunctions2}. The arrows denoted with (a)---(d) in Fig. \ref{f-eigenenergs1} correspond to the spectral densities shown in Fig. \ref{f-timevol1}, except that here we use smaller value of $N=60$ for better visibility of the spectral lines.

The gap between the two transverse red lines Fig.~\ref{f-eigenenergs1} indicate the range of eigenstates with dominant contribution to the state being squeezed, for $N=60$. The relative contributions of the eigenstates vary with $\chi_1$ since the red lines cut across the black lines that trace the eigenenergies as functions of $\chi_1$.   The histogram in Fig. \ref{f-eigenenergs1} shows the probabilities of the individual eigenstates for $\chi_1=-0.373$. The spread (twice the energy uncertainty $\Delta \epsilon$) of the histogram corresponds to separation of the two red lines mentioned above, and the correspond gap for  $N=300$ equals the width of vertical red bars in the spectral densities plotted in the insets of Fig.~\ref{f-timevol1}.

In Fig. \ref{f-Qfunctions1} we show the $Q$-functions of the Hamiltonian eigenstates in the vicinity of the peak of the energy spectrum, for a fixed value of $\chi_1=-0.373$ with all the other parameters the same as in Fig.~\ref{f-eigenenergs1}, and the fastest squeezing condition yields $u_x=0.22 N$.
As can be seen in Fig. \ref{f-Qfunctions1}, the density peak divides the spectrum into qualitatively different regions. States with energy below the peak are doubly degenerate and their  $Q$-functions form two rings encircling the poles. States with energy above the peak have  $Q$-functions following a single curve similar to a tennis-ball seam. The transition between these two regimes at the spectral peak contains pillow-shaped and X-shaped structures.

The highlighted panels in Fig. \ref{f-Qfunctions1} show eigenstates involved in the superposition being squeezed.
One can understand the squeezing process under the optimum rotation condition as gradually forming destructive interference of one diagonal pair of corners of the pillow-shaped and X-shaped structures, and constructive interference of the other pair of corners. Snapshots of the time evolution of the superposition state being squeezed is shown in  Fig. \ref{f-Qfunctions2} as the squeezing is generated, up to a time beyond when maximum squeezing is achieved.

\section{Conditions for the physical parameters}
\label{conditions}

The evolution is confined to the tangential degree of freedom if the radial and axial confinement is sufficiently tight so that occupation of higher modes of these degrees of freedom is energetically forbidden. This is achieved if
\begin{eqnarray}
&&\frac{\hbar q^2}{2mR^2}\ll \omega_{R,z},\n\\
&\equiv& \sigma_{R,z} \ll \frac{R}{q} = \frac{\lambda_a}{2\pi},
\end{eqnarray}
where $\lambda_a$ is the atomic de Broglie wavelength along the ring.

To avoid transitions to other circular modes, the corresponding couplings should be sufficiently small. In the Hamiltonian (\ref{Hama}) the terms with $u_{xy}$ couple modes $n=\pm q$ also to unwanted modes $n=\pm 3q$. The probability of such transitions would oscillate with small magnitude near zero provided that the couplings are much smaller than the energy difference of the coupled states, namely
\begin{eqnarray}
 u_{x,y} \ll \hbar\frac{(3q)^2-q^2}{2mR^2} = \frac{4q^2 \hbar}{mR^2}.
\end{eqnarray}
This condition puts a lower bound on the time during which the system can complete a $\pi$-flip between the $\pm q$ states:
\begin{eqnarray}
\tau_{\pi} = \frac{\pi}{u_{x,y}} \gg \frac{m \lambda_a^2}{16\pi \hbar}.
\end{eqnarray}
For rubidium atoms in visible light, $\tau_{\pi} \gg 20 \mu$s, so the flipping time would take at least $\sim 100$ $\mu$s.

The nonlinear term in Hamiltonian (\ref{Hama}) couples the modes $n=\pm q$ to other modes $k,l,s$ for which the condition $l+s-k =\pm q$ (homogeneous part $\chi_0$) or  $l+s-k =\pm 3q, \pm 5q$  (inhomogeneous part $\chi_1$)  is satisfied. If, say mode $+q$ is occupied with $N$ atoms and the remaining modes are empty, then the energetically closest state satisfying this condition corresponds to $N-2$ atoms in mode $q$ and one atom in each of the modes $q \pm 1$. The energy difference between these two states is $\Delta E = \hbar^2 / (mR^2)$. Thus, to avoid such transitions, the nonlinear coupling should satisfy
\begin{eqnarray}
\frac{g_{0,1}N}{16 \pi^2 \sigma_R \sigma_z R}\ll \frac{\hbar^2}{mR^2},
\end{eqnarray}
setting limits on
the shortest time during which useful squeezing can be produced
\begin{eqnarray}
\tau_{\rm sq}=\frac{1}{N \chi_{0,1}} \gg \frac{mR^2}{2 \hbar},
\end{eqnarray}
much longer than the rotational period of the superposition of two lowest-energy eigenstates of the ring trap.
Assuming rubidium in a ring trap with  $R \approx 5 \mu$ one finds $\tau_{sq}\gg 20$ ms. The process must be performed before losses become significant. With $N\approx 300$ and $\sigma_{R,z}\approx 0.3 \mu$, the peak atomic density $\approx 5.6\times 10^{13}$ cm$^{-3}$ and the dominant mechanism of losses are inelastic collisions. Recent experiments with combined magnetic and optical \cite{Bauer09} and purely optical Feshbach resonance \cite{Yan13} report suppressing inelastic loss rates to $K_{\rm in}\approx 1\times 10^{-12}$ cm$^3$/s, i.e., the requirements are just on the edge of present experimental possibilities. Full demonstration of these effects would require further suppression of $K_{\rm in}$ that can be expected, e.g., from dark-state-like interferences by means of multiphoton transitions in combined optical fields \cite{Wu}. More detailed treatment of the problem of squeezing with losses and finite temperature can be found in \cite{Losses}.

\section{State preparation and detection}
\label{stateprep}

A spin coherent state can be prepared by rotating a BEC with the chosen winding number $q$. One way to do that is via Raman transition to transfer the angular momentum from LG beams to a non-rotating BEC \cite{Ramanathan}. The resulting state is an eigenstate of $\op{J}_{z}$ with eigenvalue $N/2$. Any other spin coherent state can be prepared from this state by application of operators  $\op{J}_{x,y,z}$, by varying the parameters $u_{x,y}$ and $\omega$ to rotate the state on the Bloch sphere.

For detection $\op{J}_{x,y,z}$ , the procedure can be reversed: first the state is rotated so that the operator of interest corresponds to $\op{J}_{z}$, then a Raman process converts the $J_z=N/2$ eigenstate to a non-rotating BEC. Finally the proportion of the non-rotating atoms is measured; for the setup in Ref.~\cite{Ramanathan} this can be done by releasing the condensate and observing the size of the central hole in the interference pattern produced by the free-falling atoms. Alternately, one can measure $\op{J}_{z}$ by coupling the ring resonator to a linear atomic waveguide (formed, e.g., by a red-detuned horizontal laser beam) positioned tangentially near the ring. Atoms circulating in the opposite orientations would leak to the waveguide and propagate in opposite directions towards the waveguide ends where they can be detected.

\section{Conclusion}
\label{conclusion}

We have shown that a BEC in a toroidal trap with a circular lattice and a spatially modulated nonlinearity can be used to realize a general LMG model and comprehensive spin-squeezing procedures, including direct two-axes twisting; with independent control over all the relevant parameters. Our model should be useful for creating squeezed states for quantum measurements and testing critical phenomena in a range of physical systems described by the LMG model. It also provides a different realm of physics to be explored with the rapidly advancing technology of toroidal trapping of ultracold atoms. We conclude with some comments about our model: Unlike most schemes for bosonic pseudospin squeezing  \cite{Esteve2008,BosonicJosephson}, it can directly generate rotation about an arbitrary Bloch axis by varying  $u_{x,y}$ and $\omega$. Any relevant ratio of the LMG parameters can be achieved by co-ordinate rotation, if $\chi_1$  can access $[0,\frac{2}{3}\chi_0]$ or $[0,-\frac{2}{3}\chi_0]$. Although our examples did not use it, our model allows additional control via temporal modulation ($\omega\neq0$) of the nonlinearity.

\acknowledgments
KKD acknowledges support of the National Science Foundation under Grant No. PHY-1313871, and of a PASSHE-FPDC grant and a research grant from Kutztown University.

\appendix
\begin{widetext}

\section{Derivation of the Hamiltonian in the collective spin form}
\label{appendixA}
We start from Eq.~(\ref{Hama}) and assume that only two modes $\pm q$ are populated. We can then drop a diagonal term that only gives a constant offset. Using the definition of the non-linear coefficients $\chi_{0,1}$ the Hamiltonian takes the form
\begin{eqnarray}
 H &=&  \frac{\hbar u_x}{2}\left( \op{a}_{q}^{\dag} \op{a}_{-q} e^{i2q\omega t} + \op{a}_{-q}^{\dag}\op{a}_{q} e^{-i2q\omega t} \right) + \frac{\hbar u_y}{2i}\left( \op{a}_{q}^{\dag}\op{a}_{-q} e^{i2q\omega t} - \op{a}_{-q}^{\dag}\op{a}_{q} e^{-i2q\omega t} \right) \nonumber \\
& & + \frac{\hbar \chi_0}{2} \left( \op{a}_{q}^{\dag 2}\op{a}_{q}^{2} + 4 \op{a}_{q}^{\dag}\op{a}_{q}\op{a}_{-q}^{\dag} \op{a}_{-q} +\op{a}_{-q}^{\dag 2}\op{a}_{-q}^{2} \right)  + \frac{\hbar \chi_1}{4} \left( \op{a}_{q}^{\dag 2}\op{a}_{-q}^{2} e^{i(4q\omega t-\alpha)}  +\op{a}_{-q}^{\dag 2}\op{a}_{q}^{2} e^{-i(4q\omega t-\alpha)} \right).
\label{Hama2}
\end{eqnarray}
The time dependence of the Hamiltonian can be absorbed in the definition of the bosonic operators as
\begin{eqnarray}
 \op{a}_{q} &\equiv& \op{a} e^{iq\omega t} , \\
 \op{a}_{-q} &\equiv& \op{b} e^{-iq\omega t} .
\label{aboper}
\end{eqnarray}
Expressing the equations of motion of these operators, one finds
\begin{eqnarray}
 i \dot{\op{a}}_{q} &=& \frac{1}{\hbar}[\op{a}_{q},H] \nonumber \\
&=& \frac{u_x - iu_y}{2}\op{a}_{-q}e^{i2q\omega t}
+  \chi_0 \left(\op{a}_{q}^{\dag} \op{a}_{q}^{2} + 2\op{a}_{q} \op{a}_{-q}^{\dag}\op{a}_{-q}\right) +\frac{\chi_1}{2} \op{a}_{q}^{\dag} \op{a}_{-q}^{2} e^{i(4q\omega t-\alpha)}\nonumber \\
&=& \frac{u_x - iu_y}{2}\op{b}e^{iq\omega t}
+  \chi_0 \left(\op{a}^{\dag} \op{a}^{2} + 2\op{a} \op{b}^{\dag}\op{b}\right) e^{iq\omega t}  + \frac{\chi_1}{2} \op{a}^{\dag} \op{b}^{2} e^{iq\omega t-i\alpha}\nonumber \\
&=& i \dot{\op{a}} e^{iq\omega t} - q\omega \op{a} e^{iq\omega t},
\label{amotion}
\\
 i \dot{\op{a}}_{-q} &=& \frac{1}{\hbar}[\op{a}_{-q},H] \nonumber \\
&=& \frac{u_x + iu_y}{2}\op{a}_{q}e^{-i2q\omega t}
+  \chi_0 \left( 2\op{a}_{q}^{\dag}\op{a}_{q} \op{a}_{-q} + \op{a}_{-q}^{\dag} \op{a}_{-q}^{2} \right) +\frac{\chi_1}{2} \op{a}_{-q}^{\dag} \op{a}_{q}^{2} e^{-i(4q\omega t-\alpha)}\nonumber \\
&=& \frac{u_x + iu_y}{2}\op{a}e^{-iq\omega t}
+  \chi_0 \left(2\op{a}^{\dag} \op{a} \op{b} +  \op{b}^{\dag}\op{b}^2 \right) e^{-iq\omega t}  + \frac{\chi_1}{2} \op{b}^{\dag} \op{a}^{2} e^{-iq\omega t+i\alpha}\nonumber \\
&=& i \dot{\op{b}} e^{-iq\omega t} + q\omega \op{b} e^{-iq\omega t}.
\label{amotion2}
\end{eqnarray}
Thus, the equations of motion for operators  $\op{a}$ and  $\op{b}$ can be written as
\begin{eqnarray}
 i \dot{\op{a}} &=& q\omega \op{a} + \frac{u_x - iu_y}{2}\op{b}
+  \chi_0 \left(\op{a}^{\dag} \op{a}^{2} + 2\op{a} \op{b}^{\dag}\op{b}\right)
 + \frac{\chi_1}{2}e^{-i\alpha}\op{a}^{\dag} \op{b}^2,
\label{amotion3}
\\
 i \dot{\op{b}} &=&  -q\omega \op{b} + \frac{u_x + iu_y}{2}\op{a}
+  \chi_0 \left(2\op{a}^{\dag} \op{a} \op{b} +  \op{b}^{\dag}\op{b}^2 \right)
  + \frac{\chi_1}{2}e^{i\alpha}\op{b}^{\dag} \op{a}^2.
\label{amotion4}
\end{eqnarray}
These equations correspond to the Hamiltonian
\begin{eqnarray}
H &=& \frac{\hbar u_x}{2}\left(\op{a}^{\dag} \op{b} + \op{a}  \op{b}^{\dag}\right)
+ \frac{\hbar u_y}{2i}\left(\op{a}^{\dag} \op{b} - \op{a}  \op{b}^{\dag} \right)
 + \hbar q \omega \left(\op{a}^{\dag} \op{a} -  \op{b}^{\dag} \op{b}\right)
\nonumber  \\
& & + \frac{\hbar \chi_0}{2} \left(\op{a}^{\dag 2} \op{a}^{2} + 4
\op{a}^{\dag} \op{a}  \op{b}^{\dag} \op{b} +  \op{b}^{\dag 2} \op{b}^{2} \right)  + \frac{\hbar \chi_1}{4} \left(e^{-i\alpha}\op{a}^{\dag 2} \op{b}^{2}
+ e^{i\alpha}\op{b}^{\dag 2} \op{a}^{2} \right) ,
\label{pomham2}
\end{eqnarray}
which does not depend on time.
Using the collective spin operators $\op{J}_{x,y,z}$ of Eqs. (\ref{Jx})--(\ref{Jz}) , one can express the higher powers of the operators of Eq. (\ref{pomham2}) as
\begin{eqnarray}
\op{a}^{\dag 2}\op{a}^2 &=& \frac{N}{2}\left(\frac{N}{2} -1 \right)
+ (N-1) \op{J}_z + \op{J}_z^2, \\
\op{b}^{\dag 2}\op{b}^2 &=& \frac{N}{2}\left(\frac{N}{2} -1 \right)
- (N-1) \op{J}_z + \op{J}_z^2, \\
\op{a}^{\dag}\op{a}\op{b}^{\dag}\op{b} &=&  \left( \frac{N}{2} \right)^2 - \op{J}_z^2 , \\
\op{a}^{\dag 2}\op{b}^2+ \op{a}^2 \op{b}^{\dag 2} &=& 2 \left( \op{J}_x^2 - \op{J}_y^2 \right) ,\\
i \left( \op{a}^{\dag 2}\op{b}^2- \op{a}^2 \op{b}^{\dag 2}\right) &=& -2 \left( \op{J}_x \op{J}_y  + \op{J}_y \op{J}_x \right)   .
\end{eqnarray}
The Hamiltonian then can be written as
\begin{eqnarray}
\hat{H} = \hbar \left[u_x \op{J}_x + u_y \op{J}_y  + 2q\omega \op{J}_z - \chi_0 \op{J}_z^2  +\frac{\chi_1 \cos \alpha }{2}\left(\op{J}_x^2 - \op{J}_y^2 \right)
+\frac{\chi_1 \sin \alpha }{2}\left(\op{J}_x \op{J}_y + \op{J}_y \op{J}_x \right)
 + \chi_0 \left( \frac{3}{4}N^2 -\frac{1}{2}N \right) \right] .
\end{eqnarray}
Dropping the irrelevant last term, one obtains  Eq. (\ref{HamJ}).

\section{Squeezing parameter}
\label{appendixB}

We define the squeezing parameter $\xi^2$, following Ref. \cite{Wineland1994},  as the ratio of the smallest variance of the uncertainty ellipse
${\cal V}_{\rm min}$ on the Bloch sphere to the variance of the spin coherent state ${\cal V}_{\rm coh} = N/4$, i.e.,
\begin{eqnarray}
\xi^2 = \frac{4 {\cal V}_{\rm min}}{N}.
\end{eqnarray}
The value ${\cal V}_{\rm min}$ is determined as the smaller eigenvalue of the 2$\times$2 projection of the variance matrix to the tangential plane of the Bloch sphere with respect to the mean spin vector.
The calculation proceeds as follows.
Assume the state is characterized by the first  moments ${\cal J}_k \equiv \langle \hat J_k \rangle$ and variances $V_{kl} \equiv \frac{1}{2}\langle \hat J_k  \hat J_l + \hat J_l \hat J_k \rangle - {\cal J}_k {\cal J}_l$.  We calculate the rotated variance matrix as
\begin{eqnarray}
\tilde V = UVU^{\dag},
\label{tildeV}
\end{eqnarray}
where $V$ is matrix composed of $V_{kl}$ and
\begin{eqnarray}
U=\left(
\begin{array}{ccc}
\frac{{\cal J}_x{\cal J}_z}{|{\cal J}|\sqrt{{\cal J}_x^2 +{\cal J}_y^2}}
& \frac{{\cal J}_y{\cal J}_z}{|{\cal J}|\sqrt{{\cal J}_x^2 +{\cal J}_y^2}}
& -\frac{\sqrt{{\cal J}_x^2 +{\cal J}_y^2}}{|{\cal J}|} \\
-\frac{{\cal J}_y}{\sqrt{{\cal J}_x^2 +{\cal J}_y^2}}
& \frac{{\cal J}_x}{\sqrt{{\cal J}_x^2 +{\cal J}_y^2}} & 0 \\
\frac{{\cal J}_x}{|{\cal J}|} & \frac{{\cal J}_y}{|{\cal J}|}
& \frac{{\cal J}_z}{|{\cal J}|}
\end{array}
\right)
\end{eqnarray}
with $|{\cal J}|\equiv \sqrt{{\cal J}_x^2 +{\cal J}_y^2 + {\cal J}_z^2}$.
Note that $U$ rotates the state to the pole of the Bloch sphere, in particular
\begin{eqnarray}
U \left(
\begin{array}{c}
{\cal J}_x \\ {\cal J}_y \\ {\cal J}_z
\end{array}
\right) =
\left(
\begin{array}{c}
0 \\ 0 \\ |{\cal J}|
\end{array}
\right).
\end{eqnarray}
The value ${\cal V}_{\rm min}$ is then calculated as the smaller eigenvalue of the matrix
$V_{\rm tan}$ with
\begin{eqnarray}
V_{\rm tan} =
\left(
\begin{array}{cc}
\tilde V_{11} & \tilde V_{12} \\
\tilde V_{21} & \tilde V_{22}
\end{array}
\right) ,
\end{eqnarray}
where  $\tilde V_{kl}$ are the elements of $\tilde V$ defined in (\ref{tildeV}).

\end{widetext}

\end{document}